\documentclass[aps,prd,onecolumn,11pt,tightenlines,amsmath,amssymb,superscriptaddress,longbibliography,floatfix]{revtex4-2}
\usepackage{graphicx,bm,array,booktabs,xcolor}

\usepackage{hyperref}
\hypersetup{colorlinks=true,citecolor=black,linkcolor=black,urlcolor=black,pdftitle={Narrow-band constraints on primordial gravitational waves from CMB temperature and polarization}}
\newcommand{\dd}{\mathrm d}
\newcommand{\ii}{\mathrm i}
\newcommand{\Pt}{P_t}
\begin{document}
\raggedbottom
\title{\texorpdfstring{{Narrow-band constraints on primordial gravitational waves from CMB temperature and polarization}}{Narrow-band constraints on primordial gravitational waves from CMB temperature and polarization}}
\author{{Yuki Inoue}}
\thanks{Corresponding author: iyuki@ncu.edu.tw}
\affiliation{{Department of Physics,National Central University, Taoyuan, Taiwan}}
\affiliation{{Center for High Energy and High Field (CHiP), National Central University,  Taoyuan, Taiwan}}
\affiliation{{Institute of Physics, Academia Sinica, Taipei, Taiwan}}
\affiliation{{Institute of Particle and Nuclear Studies, High Energy Acceleration Research Organization (KEK), Tsukuba, Japan}}

\author{{Masashi Hazumi}}
\affiliation{{Department of Physics,National Central University, Taoyuan, Taiwan}}
\affiliation{{Center for High Energy and High Field (CHiP), National Central University,  Taoyuan, Taiwan}}
\affiliation{{INFN Sezione di Ferrara, Via Saragat 1, 44122 Ferrara, Italy}}

\author{{Hsiang-Yu Huang}}
\affiliation{{Department of Physics,National Central University, Taoyuan, Taiwan}}
\affiliation{{Center for High Energy and High Field (CHiP), National Central University,  Taoyuan, Taiwan}}

\begin{abstract}
{We calculate the cosmic microwave background (CMB) response to finite wavenumber bands and derive observational constraints over a broad range of scales without assuming a global shape for the primordial gravitational-wave spectrum. We generalize the semi-analytic EE and BB formalism to TT, TE, and the deflection curl, and combine it with direct CLASS calculations to evaluate 25 bands. The maximum band center, $k_c\simeq139\,\mathrm{Mpc}^{-1}$, extends the numerical analysis of Namikawa et al., whose maximum center is approximately $7.8\,\mathrm{Mpc}^{-1}$, by a factor of about 18. We retain the TT--TE--EE correlations in the public Planck, ACT, and SPT covariance matrices and combine five experiments, including BICEP/Keck and POLARBEAR. Assuming independence between experiments, the strongest combined 95\% credible upper limit on the band-averaged density is $\langle\Omega_{\rm GW}\rangle_{k,95}h^2=2.9\times10^{-16}$ at $k_c\simeq7.8\times10^{-3}\,\mathrm{Mpc}^{-1}$, a factor of about 2.6 tighter than the published BB limit of Namikawa et al. Near $k\sim1\,\mathrm{Mpc}^{-1}$, the limit reaches $1.6\times10^{-12}$ in the band centered at $k_c\simeq0.78\,\mathrm{Mpc}^{-1}$, a factor of about 52 tighter than their published TT limit. These comparisons include differences in data, likelihoods, and density-conversion conventions. Maximizing the five-experiment upper limit within a specified common-correlation family, $0\le\rho\le0.95$, gives limits 1.0--1.5 times those obtained under independence. These conditional limits quantify the temperature and polarization sensitivity to localized primordial gravitational waves up to high wavenumber.}
\end{abstract}
\maketitle
\section{Introduction}
\begingroup
The spectrum of primordial gravitational waves reflects their production in the early Universe and the subsequent cosmic expansion. Observational constraints on this spectrum therefore provide a probe of early-Universe physics~\cite{CapriniFigueroa2018}.

The direct detection of the binary black hole merger GW150914 by Advanced LIGO established gravitational-wave observation with laser interferometers~\cite{GW150914}. In addition to signals from individual sources, radiation from numerous unresolved sources and gravitational waves generated by inflation or early-Universe phase transitions can be described statistically as a stochastic gravitational-wave background (SGWB)~\cite{CapriniFigueroa2018}. We focus on the primordial component and evaluate its amplitude in finite wavenumber bands instead of imposing a single power law over a broad range of scales.

Gravitational-wave background searches employ complementary methods across different frequency ranges. The CMB probes long-wavelength primordial waves through their imprints on temperature, polarization, and photon propagation~\cite{Namikawa2019}. Pulsar timing arrays (PTAs) probe nanohertz frequencies; the NANOGrav 15-year data provide evidence for the inter-pulsar correlations expected from a gravitational-wave background~\cite{NANOGrav2023}. The planned space interferometer LISA targets primarily millihertz frequencies~\cite{LISA2017}. The proposed CHRONOS experiment combines cryogenic torsion bars with an optical speed meter to target frequencies from below a hertz to a few hertz~\cite{CHRONOSCQG2026}. At higher frequencies, LIGO--Virgo interferometer data constrain backgrounds in the tens-to-hundreds of hertz range~\cite{LVK2021}. Comparisons across these bands must account for both the production mechanism and the cosmological transfer of the primordial spectrum.

CMB temperature and polarization measurements have advanced through experiments with different sky coverage and angular resolution. Planck's multifrequency full-sky observations measured temperature and $E$-mode polarization over a broad angular range, including low multipoles~\cite{PlanckLikelihood,HiLLiPoP}. Among ground-based experiments, the Atacama Cosmology Telescope (ACT) in Chile and the South Pole Telescope (SPT) measure small-scale temperature and polarization with high sensitivity; ACT DR6 and SPT-3G D1 provide public TT, TE, and EE spectra~\cite{ACTDR6,SPT3GTnE}. BICEP/Keck searches for primordial $B$ modes with degree-scale polarization observations; we use BK18, which includes data through the 2018 observing season~\cite{BK18}. POLARBEAR, also in Chile, measures $E$- and $B$-mode polarization at subdegree scales~\cite{POLARBEAREE,POLARBEARBB}. We use the temperature and polarization spectra of these five experiments to compare and combine the constraints that different angular ranges and observables place on each primordial wavenumber band.

For short-wavelength gravitational waves, Namikawa et al.~\cite{Namikawa2019} showed that waves shorter than the scales corresponding to an experiment's angular resolution can still produce a finite CMB signal at observable multipoles. They constrained Mpc-scale gravitational waves using temperature, polarization, and lensing. Separately, Ng~\cite{Ng2022} treated the line-of-sight integral of the tensor Sachs--Wolfe effect analytically and extended constraints to shorter wavelengths using ACT and SPT temperature anisotropies. At high wavenumber, the metric evolution and projection kernels oscillate rapidly, leading to strong cancellations in the integrals. The numerical challenges discussed in both studies motivate an efficient and accurate calculation of the short-wavelength response.

The approximation study~\cite{InoueApproximation} derives the $B$-mode response using a thin recombination layer and the Wentzel--Kramers--Brillouin (WKB) approximation, and extends it to a semi-analytic calculation that retains the finite width of recombination and the relative projection phase. This method reduces the computational cost at high wavenumber and permits evaluation up to band centers $k_c\simeq140\,\mathrm{Mpc}^{-1}$. The polarization study~\cite{InoueEEBB} constrains narrow-band primordial gravitational waves with BICEP/Keck 2018 $EE+BB$ data and forecasts the sensitivity of the Simons Observatory.

The methodological contribution of the present work is to generalize the semi-analytic EE and BB formalism to TT, TE, and the deflection-curl auto-spectrum $ww$ with a common band normalization. We account for the different sources and projections, calculate the temperature, polarization, and curl responses in a unified framework, and test their accuracy against direct calculations. We also evaluate band amplitudes using public data from these five experiments, with a consistent treatment of the density conversion and inter-experiment correlations. Each primordial band is varied separately to obtain conditional upper limits from residuals relative to fixed reference models.

The paper is organized as follows. Section~\ref{sec:band-response} defines the finite-band normalization and CMB response. Section~\ref{sec:semi-analytic} derives the semi-analytic spectra, and Sec.~\ref{sec:validation} compares them with direct calculations and presents the high-wavenumber extension. Section~\ref{sec:inference} describes the data, likelihoods, and density conversion. Section~\ref{sec:results} presents the constraints and the effect of inter-experiment correlations. Section~\ref{sec:discussion} discusses their interpretation and the frequency-domain comparison, and Sec.~\ref{sec:conclusion} summarizes our conclusions. The Appendix describes the radial-kernel approximation near the turning point.
\endgroup
\section{Finite-band response}\label{sec:band-response}
{This section formulates the angular power spectra of CMB temperature, polarization, and deflection curl generated by primordial gravitational waves in a narrow wavenumber band. We first define the band width, integrated amplitude, and normalization connecting that amplitude to the observable spectra. We then express the response from metric evolution through temperature and polarization sources to projection on the sky in terms of transfer functions. The resulting band-average expression provides the common starting point for the approximations in Sec.~\ref{sec:semi-analytic} and the amplitude inference in Sec.~\ref{sec:inference}.}

We assume statistically isotropic, parity-symmetric primordial tensor perturbations, with no scalar--tensor cross-correlation. The labels $T$, $E$, and $B$ denote temperature, even-parity polarization, and odd-parity polarization, respectively; $w$ denotes the deflection curl defined below. We define the angular power spectra by $\langle a_{\ell m}^X a_{\ell' m'}^{Y*}\rangle=C_\ell^{XY}\delta_{\ell\ell'}\delta_{mm'}$, where $a_{\ell m}^X$ are the spherical-harmonic coefficients of the observed field, $m$ is the integer azimuthal index, and $\delta$ is the Kronecker delta. In the theoretical expressions below, $C_\ell$ denotes the tensor contribution. Fixed scalar and lensing contributions are included in the likelihood predictions.

We denote comoving wavenumber, scale factor, and conformal time by $k$, $a$, and $\eta$. A subscript 0 denotes the present epoch, with $a_0=1$ and the speed of light set to $c=1$. The physical and conformal Hubble rates are $H={(\partial_\eta a)/a^2}$ and $\mathcal H=aH$, respectively. Our power-spectrum convention includes both helicities and is fixed by
\begin{equation}
 C_\ell^{XY}=4\pi\int\Pt(k)\Delta_\ell^X(k)\Delta_\ell^Y(k)\dd\ln k
 \label{eq:cl-normalization}
\end{equation}
For a band $[k_-,k_+]$, we define the localized primordial power by
\begin{equation}
 \Pt(k)=\begin{cases}
 A_t/(2\Delta), & k_-\le k\le k_+,\\
 0, & \text{otherwise},
 \end{cases}
 \label{eq:band}
\end{equation}
Here $\Delta=\ln(k_+/k_-)$ is the logarithmic width, and $A_t=2\int\Pt\dd\ln k$ fixes the integrated amplitude in our convention. We define the normalized band average by $\langle f\rangle_k=\Delta^{-1}\int_{k_-}^{k_+}f\dd\ln k$ and write $L=\ell(\ell+1)$ and $Q_\ell=\sqrt{L(L-2)}$.

Substituting Eq.~\eqref{eq:band} into Eq.~\eqref{eq:cl-normalization} gives the dimensionless angular spectrum,
\begin{equation}
 D_\ell^{XY}\equiv\frac{L}{2\pi}C_\ell^{XY}
 =A_tL\langle\Delta_\ell^X\Delta_\ell^Y\rangle_k.
 \label{eq:band-spectrum}
\end{equation}
We evaluate templates at $A_{t,\mathrm{ref}}=1.26\times10^{-10}$, with the primordial amplitude excluded from $\Delta_\ell^X$. Equations below use dimensionless spectra; multiplication by $T_{\rm CMB}^2$, with $T_{\rm CMB}=2.7255\times10^6\,\mu\mathrm K$, gives TT, EE, TE, and BB in $\mu\mathrm K^2$. The curl spectrum remains dimensionless.

Throughout this work, $XY$ is restricted to TT, EE, TE, BB, and $ww$. Extension to parity-odd cross-spectra requires an explicit treatment of the helicity sum.

We use 25 bands with edges
\begin{equation}
 k_-=10^{-4+(i-1)/4},\qquad k_+=10^{-4+i/4}
 \quad[\mathrm{Mpc}^{-1}].
\end{equation}
The arithmetic band center is $k_c=(k_-+k_+)/2$, and all bands have the same width $\Delta=\tfrac14\ln10$.
For any positive $n$, the band moment is
\begin{equation}
\langle k^{-n}\rangle_k=\frac{k_-^{-n}-k_+^{-n}}{n\Delta}.
\label{eq:moment}
\end{equation}
For a fixed band shape, replacing this moment by $k_c^{-n}$ leaves a finite normalization difference. A narrow band is therefore distinct from the monochromatic limit.

\subsection{Transfer functions and radial kernels}
{We now introduce the transfer functions entering TT, EE, TE, BB, and $ww$. Our earlier studies developed the semi-analytic EE and BB formalism for finite wavenumber bands~\cite{InoueEEBB,InoueApproximation}; here we generalize it to TT, TE, and $ww$. We denote the temperature, $E$-polarization, $B$-polarization, and deflection-curl transfer functions by $\Delta_\ell^T$, $\Delta_\ell^E$, $\Delta_\ell^B$, and $\Delta_\ell^w$, respectively. TT, EE, BB, and $ww$ follow from their squares, whereas TE follows from $\Delta_\ell^T\Delta_\ell^E$, with the band average in Eq.~\eqref{eq:band-spectrum}. We express the established line-of-sight formulas~\cite{HuWhite,Namikawa} in our convention before deriving the semi-analytic responses in Sec.~\ref{sec:semi-analytic}.}

We define the comoving line-of-sight distance as $\chi=\eta_0-\eta$ and the dimensionless distance as $x=k\chi$. The optical depth, positive scattering rate, transmission, and visibility are $\tau(\eta)=\int_\eta^{\eta_0}q(\tilde\eta)\dd\tilde\eta$, $q=-{\partial_\eta\tau}=an_e\sigma_T$, $E=e^{-\tau}$, and $g=qE$, respectively. Here $n_e$ is the physical free-electron number density and $\sigma_T$ is the Thomson cross section. The function $E(\eta)$ denotes transmission, whereas a superscript $E$ labels polarization. We write $j_\ell$ for the spherical Bessel function of the first kind. {The metric transfer function $T(k,\eta)$ describes the evolution of each tensor metric mode and is normalized to unity initially outside the horizon. It is distinct from the temperature label $T$. The polarization source is defined by $p=\sqrt6\,[\Theta_2^{(2)}-\sqrt6 E_2^{(2)}]/10$, combining the tensor temperature quadrupole $\Theta_2^{(2)}$ and $E$-polarization quadrupole $E_2^{(2)}$. The superscript $(2)$ denotes the tensor component, and the quadrupoles have the same initial-amplitude normalization as $T$.} The CMB time integrals extend from the initial epoch to the present.
{We define the temperature transfer function, including metric driving and scattering, by}
\begin{equation}
 \Delta_\ell^T=\frac{Q_\ell}{4}\int\dd\eta\,[-E{\partial_\eta T}+gp]\frac{j_\ell(x)}{x^2}.
 \label{eq:los-T}
\end{equation}
The common polarization source projects onto the two parity components through their respective radial kernels:
\begin{align}
 \Delta_\ell^E&=\int\dd\eta\,gp\,r_\ell^E(x),\notag\\
 \Delta_\ell^B&=\int\dd\eta\,gp\,r_\ell^B(x).
 \label{eq:los-pol}
\end{align}
{The curl arises from the deflection accumulated as photons propagate through tensor metric perturbations from the source plane to the observer. Its transfer integral therefore differs from the polarization integral weighted by the visibility near recombination. We retain the metric evolution and radial kernel along the full photon path, rather than localizing the integral to the recombination layer:}
\begin{equation}
 \Delta_\ell^w=\frac{Q_\ell}{2}\int_0^{\chi_*}\dd\chi\,kT(k,\eta_0-\chi)\frac{j_\ell(x)}{x^2}.
 \label{eq:los-w}
\end{equation}
{The polarization radial kernels are~\cite{HuWhite}}
\begin{align}
 r_\ell^E(x)&=\frac14\left[{\partial_x^2 j_\ell(x)}+\frac4x{\partial_x j_\ell(x)}+\left(\frac2{x^2}-1\right)j_\ell\right],\notag\\
 r_\ell^B(x)&=\frac12\left[{\partial_x j_\ell(x)}+\frac2xj_\ell\right].
 \label{eq:radial-pol}
\end{align}
{The operators $\partial_x$ and $\partial_\eta$ denote derivatives with respect to $x$ and conformal time $\eta$, respectively. For functions of several arguments, the other independent variables are held fixed.} TE is obtained by integrating $\Delta_\ell^T\Delta_\ell^E$ over the band, with no additional source.

{We define the deflection curl in Eq.~\eqref{eq:los-w} by $w=-\nabla^2\varpi$.} Here $\varpi$ is the curl potential in $\alpha_a=\nabla_a\phi+\epsilon_a{}^b\nabla_b\varpi$~\cite{Namikawa}, $\phi$ is the gradient potential, $\epsilon_a{}^b$ is the antisymmetric tensor, and $\nabla_a$ is the covariant derivative on the sphere. Thus $w_{\ell m}=L\varpi_{\ell m}$ and $C_\ell^{ww}=L^2C_\ell^{\varpi\varpi}$. For the rotation angle $\omega=w/2$, $C_\ell^{\omega\omega}=C_\ell^{ww}/4$.

We define $\eta_*$ as the time of the recombination visibility peak and $\chi_*=\eta_0-\eta_*$ as its comoving distance. {In the theoretical curl spectrum, we approximate the recombination layer by a sharp source plane at this epoch.}

Since $j_\ell(x)/x^2\sim x^{\ell-2}/(2\ell+1)!!$, the integration kernel is regular at the origin for $\ell\ge2$. Applying the large-$x$ form $\sin x/x$ in this region produces an apparent divergence outside the domain of the approximation.

\section{Semi-analytic spectra}\label{sec:semi-analytic}\subsection{Metric evolution and photon response}
For modes that enter the horizon during radiation domination, the short-wavelength metric transfer is approximated by~\cite{Weinberg,Dicus}
\begin{equation}
 T(k,\eta)\simeq\frac{\mathcal B}{k}\operatorname{Re}[U(\eta)e^{\ii(k\eta+\varphi)}].
 \label{eq:metric}
\end{equation}
Here {$U(\eta)=1/a(\eta)$ describes the dilution of the wave amplitude by cosmic expansion as a function of conformal time; its argument is suppressed below where convenient.} The radiation-era normalization is $\mathcal B=D_\nu H_0\sqrt{\Omega_r}$, where $\Omega_r$ is the radiation density fraction and $D_\nu$ is the neutrino amplitude-damping factor. The phase $\varphi(k)$ is independent of time. This expression applies after horizon entry for $k$ well above the wavenumber at matter--radiation equality. Matching to the radiation-era solution $T=\sin(k\eta)/(k\eta)$ sets the normalization before neutrino damping.
The choice of sine or cosine is absorbed into the common phase $\varphi$. Since $D_\nu$ acts on the amplitude, the power contains $D_\nu^2$.

We define the cosine of the angle between the photon direction and the wavevector as $\mu=\hat{\bm k}\cdot\hat{\bm n}$. The photon temperature and polarization perturbations $F(k,\mu,\eta)$ and $G(k,\mu,\eta)$ have the tensor azimuthal dependence factored out. Their convention is specified by the source equations
\begin{align}
 (\partial_\eta+\ii k\mu+q)F&={\partial_\eta T}-qp,\notag\\
 (\partial_\eta+\ii k\mu+q)G&=qp.
 \label{eq:photons}
\end{align}
The metric time derivative drives the temperature perturbation, while scattering couples both photon variables to the polarization source. This source is their weighted angular moment,
\begin{equation*}
 p=\frac12\int_{-1}^1[-w_FF+w_GG]\dd\mu.
\end{equation*}
The weights describe the angular dependence of Thomson scattering for tensor perturbations:
\begin{align}
 w_F&=\frac3{16}(1-\mu^2)^2,\notag\\
 w_G&=\frac3{16}(1+6\mu^2+\mu^4).
 \label{eq:weights}
\end{align}
With $\delta=q/k$ and $j_{F,G}=\tfrac12\int_{-1}^1w_{F,G}/[\delta+\ii(1+\mu)]\dd\mu$, the local stationary response is
\begin{equation}
 R(\delta)=-\frac{\ii j_F}{1-\delta(j_F+j_G)}.
 \label{eq:stationary-response}
\end{equation}
As $\delta\to0$, the angular integrals satisfy $j_F(0)=1/(8\ii)$ and $\delta j_G\to0$. The source response therefore approaches $R\to-1/8$, which fixes the coefficient in the coherent polarization limit.
After factoring out $e^{\ii(k\eta+\varphi)}$, $R$ is the complex amplitude ratio of the polarization source to $U$. The angular response integrals $j_F,j_G$ are distinct from the spherical Bessel function $j_\ell$.

We factor out $(\mathcal B/k)e^{\ii(k\eta+\varphi)}$ and expand the remaining envelopes in time derivatives. We write $d=\ii kU+{\partial_\eta U}$, $d_\mu=q+\ii k(1+\mu)$, $J_{F,G}=j_{F,G}/k$, and $M=1-q(J_F+J_G)$.
The leading envelopes are $p_0=-dJ_F/M$, $F_0=(d-qp_0)/d_\mu$, and $G_0=qp_0/d_\mu$. Their time dependence generates the first-order correction. Expanding the inverse operator $(d_\mu+\partial_\eta)^{-1}$ gives
\begin{equation*}
 p_1=\frac1{2M}\int_{-1}^1\frac{w_F{\partial_\eta F_0}-w_G{\partial_\eta G_0}}{d_\mu}\dd\mu.
\end{equation*}
Combining the leading source and its first-order correction, we define the response normalized by the metric envelope as
\begin{equation}
 R_{\rm eff}=\frac{p_0+p_1}{U}.
 \label{eq:response}
\end{equation}
Restoring the common oscillatory factor gives the real polarization source,
\begin{equation}
p(k,\eta)\simeq\frac{\mathcal B}{k}\operatorname{Re}[UR_{\rm eff}e^{\ii(k\eta+\varphi)}]
\end{equation}
This expansion has a restricted domain at low $k$ and near the resonant endpoint, where uniform accuracy requires additional treatment.

\subsection{Projected amplitudes and TE}\label{sec:complex}
We write $f_T=-E(\ii kU+{\partial_\eta U})+gUR_{\rm eff}$ and evaluate all time-dependent quantities at $\eta=\eta_0-\chi$. We then define the following projected amplitudes.
The temperature amplitude retains both metric driving and scattering through $f_T$:
\begin{equation}
 \mathcal H_T=\frac{Q_\ell}{4}\int_0^{\chi_{\max}}f_T e^{-\ii x}\frac{j_\ell(x)}{x^2}\dd\chi.
 \label{eq:H-T}
\end{equation}
For polarization, the common source is projected through the $E$- and $B$-mode radial kernels:
\begin{align}
 \mathcal H_E&=\int_0^{\chi_{\max}}gUR_{\rm eff} e^{-\ii x}r_\ell^E(x)\dd\chi,\notag\\
 \mathcal H_B&=\int_0^{\chi_{\max}}gUR_{\rm eff} e^{-\ii x}r_\ell^B(x)\dd\chi.
 \label{eq:H-pol}
\end{align}
The curl amplitude is given by the metric envelope integrated to the source plane:
\begin{equation}
 \mathcal H_w=\frac{kQ_\ell}{2}\int_0^{\chi_*}Ue^{-\ii x}\frac{j_\ell(x)}{x^2}\dd\chi.
 \label{eq:H}
\end{equation}
The upper limit $\chi_{\max}$ corresponds to a sufficiently opaque epoch, so the WKB metric is used only after this epoch rather than extrapolated to $\eta=0$. These \textbf{semi-analytic integral representations} retain the full Bessel kernels and are applicable at high multipoles, subject to the high-$k$ approximation for the source.

We describe all components by complex amplitudes $\mathcal H_X(k,\ell)$ in a common phase convention. Where band-phase averaging is valid, the spectra are
\begin{equation}
D_\ell^{XY}\simeq\frac{A_tL\mathcal B^2}{2}
\left\langle\frac{\operatorname{Re}(\mathcal H_X\mathcal H_Y^*)}{k^2}\right\rangle_k.
\label{eq:master}
\end{equation}
Here $*$ denotes complex conjugation and $\operatorname{Re}$ the real part.

Defining $\theta=k\eta_0+\varphi$, we can write $\Delta_\ell^X\simeq(\mathcal B/k)\operatorname{Re}(e^{\ii\theta}\mathcal H_X)$. The product identity gives
\begin{equation}
\Delta_\ell^X\Delta_\ell^Y\simeq\frac{\mathcal B^2}{2k^2}
\left[\operatorname{Re}(\mathcal H_X\mathcal H_Y^*)+
\operatorname{Re}(e^{2\ii\theta}\mathcal H_X\mathcal H_Y)\right].
\end{equation}
Equation~\eqref{eq:master} follows when the band integral of the second term is sufficiently small. This omission requires a check that includes the phase variation of the source as well as the explicit wavenumber dependence.

In particular, TE is
\begin{equation}
D_\ell^{TE}\simeq\frac{A_tL\mathcal B^2}{2}
\left\langle k^{-2}|\mathcal H_T||\mathcal H_E|
\cos[\arg\mathcal H_T-\arg\mathcal H_E]\right\rangle_k.
\end{equation}
The sign of $D^{TE}$ therefore depends on the relative phase and requires more information than $\sqrt{D^{TT}D^{EE}}$. With a common positive band weight, the Cauchy--Schwarz inequality gives
\begin{equation}
|D_\ell^{TE}|\le\sqrt{D_\ell^{TT}D_\ell^{EE}}.
\end{equation}
Near a zero of TE, its relative error diverges. We therefore use a normalization such as $|\delta D^{TE}|/\sqrt{D^{TT}D^{EE}}$ for numerical validation.

\subsection{WKB projection}
The function $u=xj_\ell$ satisfies ${\partial_x^2u+(1-L/x^2)u=0}$. The WKB projection applies in the oscillatory domain sufficiently far from the turning point $x=\sqrt L$.
We define the local radial wavenumber and phase by
\begin{align}
 \mu_\ell&=\sqrt{1-L/x^2},\notag\\
 \Phi_\ell&=\sqrt{x^2-L}-\sqrt L\arccos(\sqrt L/x)
 +\frac\pi2(\sqrt L-\ell).
 \label{eq:radial-phase}
\end{align}
These quantities give $j_\ell\simeq\sin\Phi_\ell/(x\sqrt{\mu_\ell})$. Retaining phase derivatives in Eq.~\eqref{eq:radial-pol} yields $r_\ell^E\simeq-\mathcal A_E\sin\Phi_\ell$ and $r_\ell^B\simeq\mathcal A_B\cos\Phi_\ell$, with amplitudes
\begin{equation}
 \mathcal A_E=\frac{1+\mu_\ell^2}{4x\sqrt{\mu_\ell}},\qquad
 \mathcal A_B=\frac{\sqrt{\mu_\ell}}{2x}.
 \label{eq:radial-amplitudes}
\end{equation}
The dimensionless quantity $\mu_\ell(x)$ is the local radial wavenumber of the projection kernel, distinct from the photon-direction cosine $\mu$. The explicit phase constant ensures $\Phi_\ell\to x-\pi\ell/2$. We retain this relative phase when matching to the exact solution in the near region.

We define the difference phase as $\Psi_\ell=\Phi_\ell-x$. In a far region $\mathcal D$ where the rapidly varying contribution $e^{-\ii(\Phi_\ell+x)}$ is negligible, we introduce
\begin{align}
 \mathcal J_T&=\int_{\mathcal D}\frac{f_T e^{\ii\Psi_\ell}}{x^3\sqrt{\mu_\ell}}\dd\chi,\notag\\
 \mathcal J_w&=\int_{\mathcal D}\frac{U e^{\ii\Psi_\ell}}{x^3\sqrt{\mu_\ell}}\dd\chi.
\end{align}
These two integrals share the metric projection kernel but have different sources. The finite-width polarization responses are
\begin{equation}
 \mathcal J_X=\int_{\mathcal D}gUR_{\rm eff}\mathcal A_Xe^{\ii\Psi_\ell}\dd\chi,
 \qquad X=E,B.
 \label{eq:J-pol}
\end{equation}
Using $e^{-\ii x}\sin\Phi_\ell\to e^{\ii\Psi_\ell}/(2\ii)$ and $e^{-\ii x}\cos\Phi_\ell\to e^{\ii\Psi_\ell}/2$, we obtain
\begin{equation}
 \mathcal H_T^{\rm far}\simeq-\frac{\ii Q_\ell}{8}\mathcal J_T,\qquad
 \mathcal H_w^{\rm far}\simeq-\frac{\ii kQ_\ell}{4}\mathcal J_w.
\end{equation}
The sine and cosine polarization kernels contribute different complex phases:
\begin{equation}
 \mathcal H_E^{\rm far}\simeq\frac\ii2\mathcal J_E,\qquad
 \mathcal H_B^{\rm far}\simeq\frac12\mathcal J_B.
\end{equation}
These relative phases are retained in the cross-spectrum.
\textbf{When the far-region contribution dominates}, Eq.~\eqref{eq:master} gives the following spectra.
\begin{subequations}\label{eq:high}
Squaring the far-region temperature amplitude gives
\begin{equation}
 D_\ell^{TT}\simeq\frac{A_tL\mathcal B^2Q_\ell^2}{128}\langle k^{-2}|\mathcal J_T|^2\rangle_k.
\end{equation}
The $E$- and $B$-mode auto-spectra share a common prefactor, while their radial amplitudes distinguish the finite-width responses:
\begin{equation}
 D_\ell^{EE,BB}\simeq\frac{A_tL\mathcal B^2}{8}\langle k^{-2}|\mathcal J_{E,B}|^2\rangle_k.
\end{equation}
The relative phase of temperature and $E$ polarization gives
\begin{equation}
 D_\ell^{TE}\simeq-\frac{A_tL\mathcal B^2Q_\ell}{32}\langle k^{-2}\operatorname{Re}(\mathcal J_T\mathcal J_E^*)\rangle_k.
\end{equation}
The additional factor of $k$ in the curl amplitude cancels the metric factor $k^{-2}$ in the power:
\begin{equation}
 D_\ell^{ww}\simeq\frac{A_tL\mathcal B^2Q_\ell^2}{32}\langle|\mathcal J_w|^2\rangle_k.
\end{equation}
\end{subequations}
The explicit minus sign in the TE expression alone does not determine its sign; the real part of the complex product must also be evaluated.

\subsection{Finite-width effects and limiting forms}
\subsubsection{Polarization phase coherence}
Even when $x\gg\sqrt L$, the phase can vary appreciably across the temporal width of the source. The large-$x$ expansion leaves the difference phase
\begin{equation}
 \Psi_\ell=-\frac{\pi\ell}{2}+\frac{L}{2k\chi}+O(L^2/x^3).
\end{equation}
Its variation with scattering time controls interference across the source layer:
\begin{equation}
 \frac{\dd\Psi_\ell}{\dd\eta}=k(1-\mu_\ell)\simeq\frac{L}{2k\chi^2}.
 \label{eq:mismatch}
\end{equation}
Thus departures from $D^{EE,BB}\propto L$ occur as $L\sigma_\eta/(2k\chi_*^2)$ approaches unity, where $\sigma_\eta$ is the conformal-time width of the source.

For a Gaussian source of width $\sigma_\eta$, with a slowly varying radial amplitude and a constant complex source phase, the power acquires the suppression factor $\exp[-\sigma_\eta^2 k^2(1-\mu_{\ell,*})^2]$. The characteristic turnover in this example is $\ell_{\rm coh}\sim(2k\chi_*^2/\sigma_\eta)^{1/2}$. Our calculation uses the visibility and complex response directly, rather than this Gaussian approximation.

\subsubsection{Matching to the near region}
For TT and $ww$, contributions near $\chi\to0$ and the turning point can be important. We then form
\begin{equation}
\mathcal H_X=\mathcal H_X^{\rm near,exact}+\mathcal H_X^{\rm far,WKB}
\end{equation}
\textbf{at the amplitude level} before using Eq.~\eqref{eq:master}, thereby retaining interference between the two regions. The near region extends from the observer to beyond the turning point and uses the full Bessel kernels.

We retain the rapidly oscillating sum phase until its contribution becomes negligible. Near $x=\sqrt L$, a uniform Airy kernel replaces the singular WKB amplitude. {The explicit kernel is given in Appendix~\ref{app:radial}.} The integral representation in Sec.~\ref{sec:complex} preserves a common phase between the near and far regions.

\subsubsection{Coherent polarization limit}
The low-multipole limit here refers to a high-wavenumber band satisfying $\ell\ll k\chi_*$ and $L\sigma_\eta/(2k\chi_*^2)\ll1$, rather than to small $k$. We use $R\to-1/8$, $r_\ell^E\simeq-\sin(x-\pi\ell/2)/(2x)$, and $r_\ell^B\simeq\cos(x-\pi\ell/2)/(2x)$. Since $\eta+\chi=\eta_0$, the product of source and projection contains a term that is nonoscillatory in time. Writing $I_{\rm rec}=\int_{\rm rec}g/(a\chi)\dd\eta$, we obtain
\begin{equation}
 \mathcal J_E\simeq\mathcal J_B\simeq-\frac{I_{\rm rec}}{16k}\,e^{-\ii\pi\ell/2}.
\end{equation}
The common phase cancels in the auto-spectra. Substitution into Eq.~\eqref{eq:high}, together with Eq.~\eqref{eq:moment}, gives
\begin{equation}
D_\ell^{EE}\simeq D_\ell^{BB}\simeq
\frac{A_tL\mathcal B^2I_{\rm rec}^2}{8192\Delta}(k_-^{-4}-k_+^{-4}).\label{eq:polarization-closed}
\end{equation}
The scaling $D_\ell\propto L$ corresponds to $C_\ell\simeq\mathrm{const.}$

\subsubsection{Temperature and curl near the observer}
In the local limit $U\simeq E\simeq1$, with $gR$ and ${\partial_\eta U}$ neglected,
the radial dependence reduces to a single integral:
\begin{equation}
 J_\ell=\int_0^\infty e^{-\ii x}\frac{j_\ell(x)}{x^2}\dd x
 =\frac{2\ii^{1-\ell}}{Q_\ell^2}.
\end{equation}
The temperature and curl amplitudes become $\mathcal H_T\simeq-\ii Q_\ell J_\ell/4$ and $\mathcal H_w\simeq Q_\ell J_\ell/2$, respectively.
{In this local approximation, the magnitude of the curl amplitude is twice that of temperature. Substitution into Eq.~\eqref{eq:master} gives the dimensionless auto-spectra,}
\begin{equation}
 {D_\ell^{ww}=4D_\ell^{TT}\simeq\frac{A_t\mathcal B^2}{4\Delta(L-2)}\left(k_-^{-2}-k_+^{-2}\right).}
\end{equation}
{The equality holds in the local approximation above, with dimensionless TT power. We have substituted the band average from Eq.~\eqref{eq:moment}. At fixed logarithmic band width, the polarization result in Eq.~\eqref{eq:polarization-closed} scales as $D_\ell^{EE},D_\ell^{BB}\propto Lk_c^{-4}$, whereas the local temperature and curl limits scale as $D_\ell^{TT},D_\ell^{ww}\propto(L-2)^{-1}k_c^{-2}$.} This local formula is inapplicable when contributions from the recombination end dominate, even at low multipoles.

\section{Numerical validation}\label{sec:validation}
{We assess the accuracy of the semi-analytic formulas by comparison with direct CLASS calculations, focusing on high wavenumbers where the numerical integrations are particularly costly. We also use the approximation to extrapolate the narrow-band spectra to still higher wavenumbers. Table~\ref{tab:parameters} summarizes the adopted parameters and corresponding background quantities. The direct and semi-analytic calculations use the same background, recombination history, and primordial amplitude.}
\begin{table}[tb]

\caption{Adopted parameters and background quantities. Background values are rounded to four or five significant figures. The conformal times $\eta_0$ and $\eta_*$ correspond to the present epoch and the visibility peak for the chosen background and recombination history. In the transfer functions we use $c=1$ and $H_0/c=2.2469\times10^{-4}\,\mathrm{Mpc}^{-1}$.}
\label{tab:parameters}
\begin{ruledtabular}
\begin{tabular}{lr}
Quantity & Adopted value\\
$H_0$ [$\mathrm{km\,s^{-1}\,Mpc^{-1}}$] & $67.36$\\
$\Omega_r$ & $9.2181\times10^{-5}$\\
$D_\nu$ & $0.8016$\\
$T_{\rm CMB}$ [K] & $2.7255$\\
$A_{t,\mathrm{ref}}$ & $1.26\times10^{-10}$\\
$\eta_0$ [Mpc] & $1.4175\times10^4$\\
$\eta_*$ [Mpc] & $280.71$\\
\end{tabular}
\end{ruledtabular}
\end{table}
{The semi-analytic polarization response describes recombination. Direct calculations provide the low-wavenumber spectra, including the large-angle polarization generated by reionization.}

\subsection{{Treatment of TE and comparison with CLASS}}
\begin{figure}[tb]
\centering\includegraphics[width=.98\textwidth]{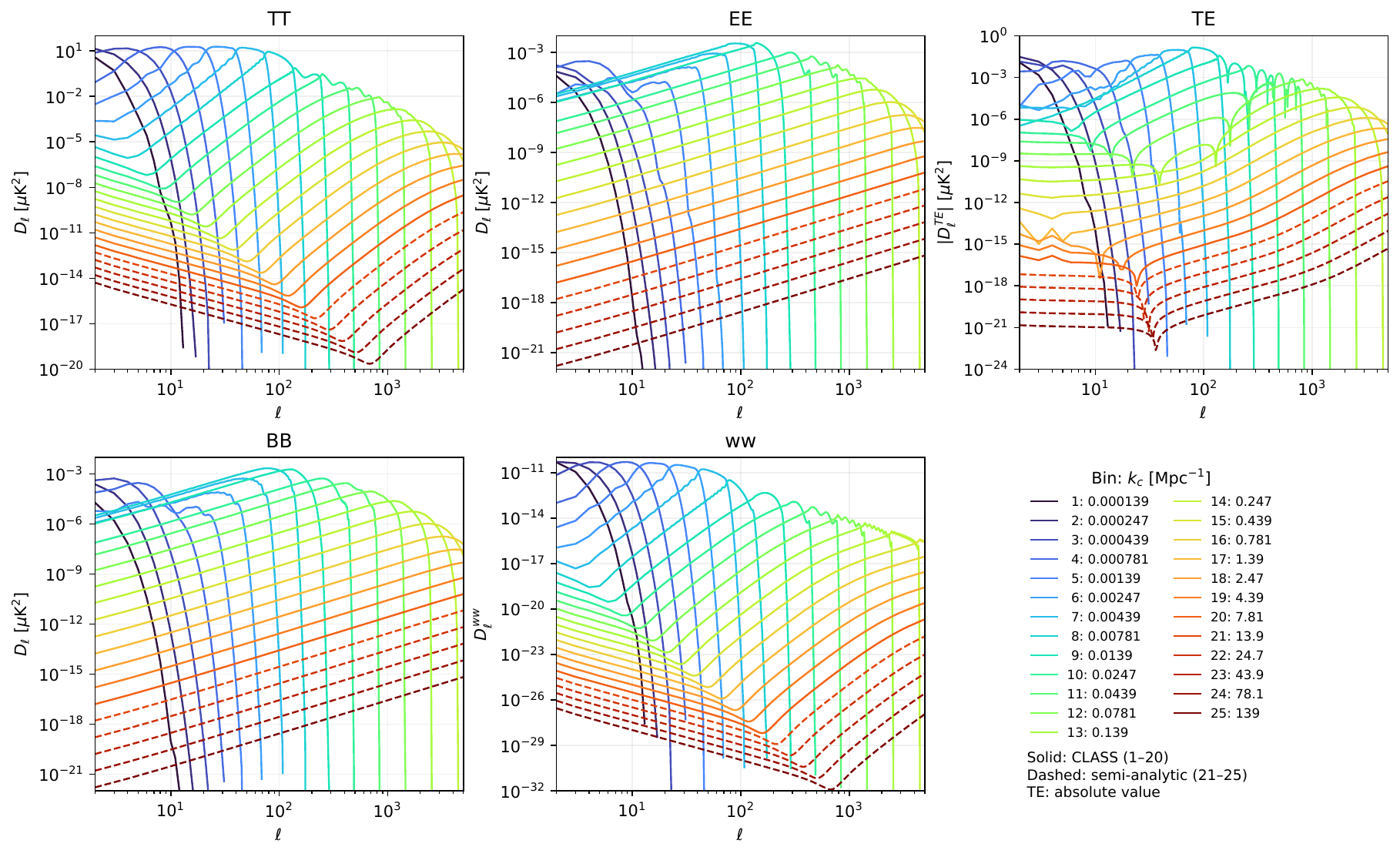}
\caption{Finite-band responses at $A_{t,\mathrm{ref}}=1.26\times10^{-10}$. Solid curves show the completed CLASS spectra for bins 1--20, and dashed curves show the semi-analytic extension to bins 21--25. Colors indicate the band center. CMB spectra are in $\mu\mathrm K^2$, while curl is dimensionless. TE is shown in absolute value; its signed high-wavenumber response is shown in Fig.~\ref{fig:class}. {The vertical ranges emphasize the main response of each component and the shape of the high-wavenumber bands.}}\label{fig:all-class}\label{fig:library}
\end{figure}

\begin{figure}[tb]
\centering\includegraphics[width=.85\textwidth]{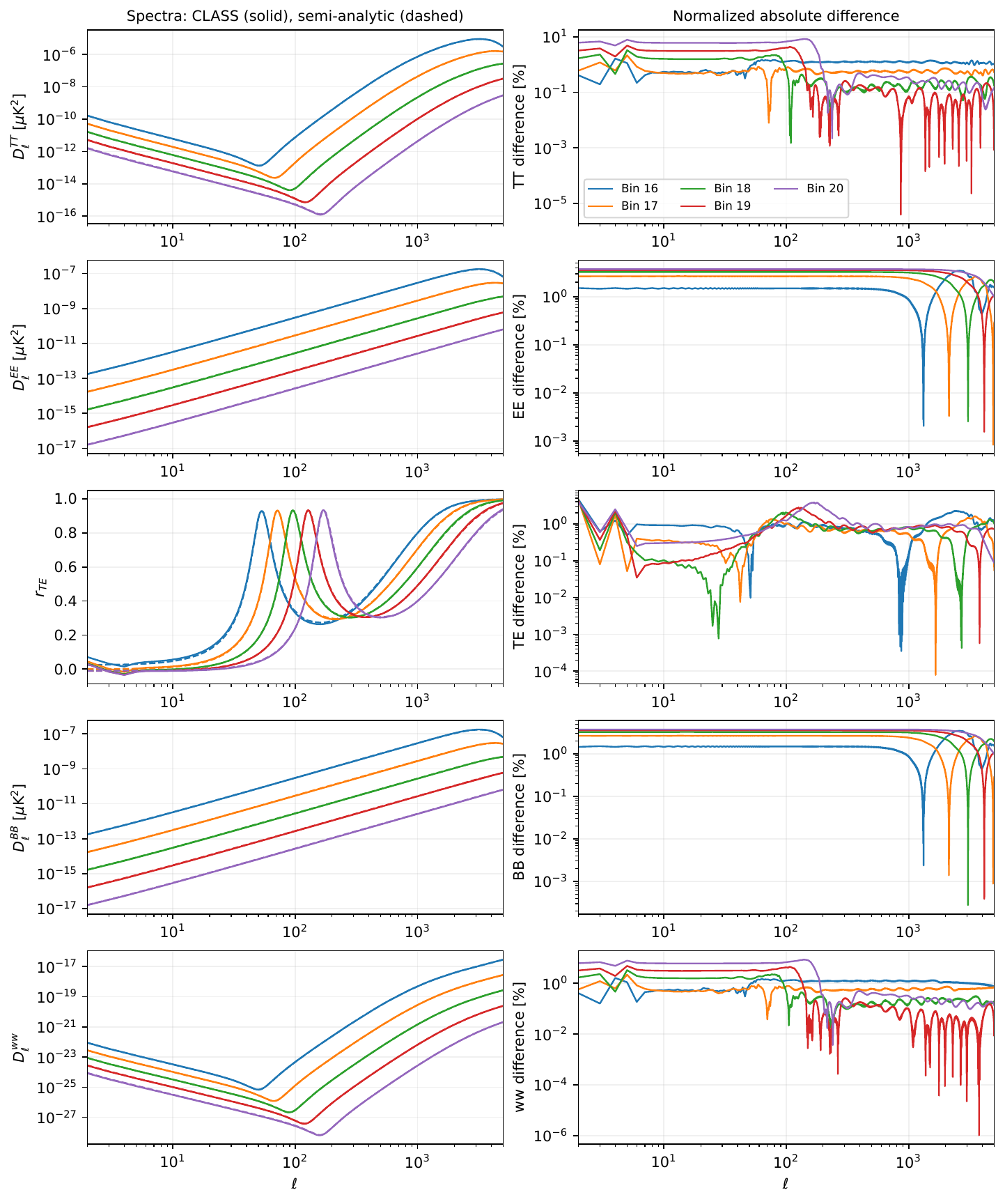}
\caption{Validation for bins 16--20. The left column compares CLASS (solid) with semi-analytic spectra (dashed) at the same primordial amplitude; TE is displayed as $r_{TE}=D^{TE}/\sqrt{D^{TT}D^{EE}}$. The right column shows absolute differences normalized by the CLASS auto-power, or by $\sqrt{D^{TT}_{\rm CLASS}D^{EE}_{\rm CLASS}}$ for TE. The denominator floor is $10^{-12}$ times the maximum of the normalization scale. No amplitude fitting is performed.}\label{fig:class}\label{fig:errors}
\end{figure}

\begingroup
TE is the cross-spectrum of temperature and $E$ polarization, and its sign depends on the relative phase of their transfer functions. A logarithmic display or relative difference alone therefore gives an incomplete account of its behavior and accuracy. We display the signed response using the correlation coefficient $r_{TE}=D_\ell^{TE}/\sqrt{D_\ell^{TT}D_\ell^{EE}}$. Where the auto-spectra are positive, $|r_{TE}|\le1$; its positive and negative signs correspond to correlation and anticorrelation between temperature and polarization. To assess accuracy near TE zero crossings, we normalize differences by $\sqrt{D_\ell^{TT}D_\ell^{EE}}$ rather than $|D_\ell^{TE}|$.

Figure~\ref{fig:all-class} compares the TT, EE, TE, BB, and $ww$ responses of 25 wavenumber bands with the same integrated amplitude $A_t$. Colors indicate the band center $k_c$, solid curves show direct CLASS calculations for bins 1--20, and dashed curves show the semi-analytic results for bins 21--25. With a common primordial amplitude, differences in curve height reflect the response of each observable. In particular, polarization retains a low-multipole response even at high wavenumber, approaching $D_\ell\propto\ell(\ell+1)$ in Eq.~\eqref{eq:polarization-closed}. Temperature and curl also receive contributions along the photon path and have a different multipole dependence. TE is shown in absolute value to accommodate its wide dynamic range; its troughs can therefore include sign changes. Figure~\ref{fig:class} provides the signed comparison.
\endgroup

\begingroup
Figure~\ref{fig:class} tests the semi-analytic calculation against independently refined direct calculations with modified CLASS~\cite{CLASS}. The rows correspond to TT, EE, TE, BB, and $ww$. The left column compares CLASS (solid) and semi-analytic (dashed) spectra for the same primordial amplitude and background, without amplitude adjustment. Displaying TE as $r_{TE}$ reveals its sign and sign changes, which are obscured in Fig.~\ref{fig:all-class}.

The right column shows the absolute differences between the two calculations. Auto-spectra are normalized by the corresponding CLASS power, and TE by the CLASS value of $\sqrt{D_\ell^{TT}D_\ell^{EE}}$.

Table~\ref{tab:comparison} gives the maximum normalized differences over $2\le\ell\le5000$ in Fig.~\ref{fig:class}. EE and BB agree within 3.8\% for bins 16--20, with 3.7\% for both in bin 20. The maximum normalized TE differences are 3.9--4.7\%, measured relative to the auto-spectrum scale above rather than to TE itself. For TT and $ww$, the differences increase toward higher wavenumber, reaching 8.3\% and 8.5\% in bin 20. This comparison distinguishes the accuracy of the polarization calculation from that of its generalization to temperature and curl.

\endgroup

\begin{table}[tb]

\caption{Maximum normalized differences (\%) between the semi-analytic calculation and the completed CLASS reference over $2\le\ell\le5000$. The denominator floor is $10^{-12}$ times the maximum of each normalization scale. TE uses $\sqrt{D^{TT}D^{EE}}$. These are approximation differences, distinct from numerical refinement differences within CLASS.}
\label{tab:comparison}
\begin{ruledtabular}
\begin{tabular}{crrrrr}
Bin & TT & EE & TE & BB & $ww$\\
16 & 1.6 & 3.5 & 4.7 & 3.5 & 1.6\\
17 & 2.2 & 2.7 & 4.5 & 2.7 & 2.2\\
18 & 3.3 & 3.3 & 4.3 & 3.3 & 3.3\\
19 & 4.8 & 3.6 & 4.2 & 3.6 & 4.8\\
20 & 8.3 & 3.7 & 3.9 & 3.7 & 8.5\\
\end{tabular}
\end{ruledtabular}
\end{table}

{The high-wavenumber extension in Fig.~\ref{fig:all-class} reaches $k_c\simeq139\,\mathrm{Mpc}^{-1}$, with band edges $k_-=100\,\mathrm{Mpc}^{-1}$ and $k_+\simeq178\,\mathrm{Mpc}^{-1}$. Bins 21--25 are semi-analytic predictions beyond the range directly compared with CLASS in Fig.~\ref{fig:class}.}

\section{Data and inference}\label{sec:inference}
\subsection{Data selection and fixed parameters}
For the primary CMB, we use BK18 EE and BB~\cite{BK18}, ACT DR6 TT, TE, and EE~\cite{ACTDR6}, SPT-3G D1 TT, TE, and EE~\cite{SPT3GTnE}, and Planck high-multipole HiLLiPoP~\cite{HiLLiPoP}, low-multipole LoLLiPoP EE~\cite{LoLLiPoP}, and Commander TT~\cite{PlanckLikelihood}. Commander is the 2018 release, so this combination includes both 2018 and PR4 products. {We also analyze POLARBEAR EE~\cite{POLARBEAREE} and BB~\cite{POLARBEARBB} separately as complementary polarization data.} We exclude SPT BB and the parity-odd TB and EB spectra. {Observational constraints are evaluated for temperature and polarization, while curl is treated at the level of theoretical spectra.} Individual-experiment limits are the primary results; Sec.~\ref{sec:cross-experiment} separately examines how assumptions about inter-experiment correlations affect combined limits.

We retain the EE--BB cross-covariance supplied with BK18 and remove observed matrix elements corresponding to EB. The selected vector has 1188 elements, with a maximum absolute EE--BB correlation coefficient of approximately {0.14}, so EE and BB are treated jointly. We also use the public TT--TE--EE covariance matrices for ACT and SPT. For Planck, we assume factorization between low-$T$ and low-$E$ likelihoods and between the low- and high-multipole blocks. This is an approximation rather than a measurement of vanishing cross-covariance.

{For the Gaussian residual analyses, we subtract a reference model from the measured spectra while retaining the public window functions and covariance. The reference background cosmology, foregrounds, and calibration are fixed. We use the public best fit for ACT, public reference foreground and calibration values for SPT and high-multipole Planck, and the distributed fiducial bandpowers for BK18. This common prescription concerns the residual-analysis method; each experiment retains its own reference spectrum. The resulting limits are conditional on these fixed models, with foreground and calibration parameters held fixed. For BK18 we use a Gaussian approximation that retains the EE--BB covariance.}

{{The fixed scalar reference for SPT and Planck uses the Planck reference $\Lambda$CDM cosmology~\cite{PlanckParameters}, with neutrinos approximated as massless; the corresponding lensed spectra are calculated with CAMB~\cite{CAMB}.} The ACT scalar reference uses its public best-fit cosmology. The added tensor response and density conversion use a common background across experiments, with only the band amplitude varied. At low Planck multipoles, the fixed reference plus the tensor signal is passed to the public non-Gaussian likelihood, preserving its likelihood shape.}

{For POLARBEAR EE, we use the 25 released bandpowers over $500\le\ell<3000$, their windows, and covariance. With the background cosmology fixed, we marginalize over foreground amplitudes and calibration following the public likelihood, including beam uncertainty. For POLARBEAR BB, we use the four bands over $500\le\ell<2100$ released in 2014 and project the theory with the discrete sum specified by the released windows. The BB analysis uses the lensed component as the reference, fixes the foreground amplitude to zero and calibration to unity, and adopts a Gaussian approximation with diagonal covariance. Figure~\ref{fig:limits} shows the upper limit from the product of the EE and BB likelihoods under an independence approximation; Sec.~\ref{sec:cross-experiment} considers its combination with other experiments.}

\subsection{Window functions and amplitude posterior}
We allow only the target band $i$ to be nonzero and define its amplitude relative to the reference as $a_i=A_{t,i}/A_{t,\mathrm{ref}}$. The predicted data vector is
\begin{equation}
\bm m_i(a_i)=\bm m_0+a_i\bm s_i.
\end{equation}
The template vector is obtained by projecting the narrow-band spectrum through the public windows:
\begin{equation}
 s_{b,i}=\sum_\ell W_{b\ell}D_{\ell,i}^{\rm NB}.
\end{equation}
Here $b$ jointly indexes observational bands and frequency--spectrum combinations, $W_{b\ell}$ is the public window function, and $\bm s_i$ is the response vector at the reference amplitude. The fixed prediction before adding the tensor signal is $\bm m_0$, the observed vector is $\bm d$, and $\bm\Sigma$ is the covariance in the same ordering. The superscript NB denotes a narrow band.

Writing the log likelihood as $\ln\mathcal L$, the Gaussian likelihood with fixed covariance is
\begin{equation}
 -2\ln\mathcal L=(\bm d-\bm m_i)^T\bm\Sigma^{-1}(\bm d-\bm m_i)+\mathrm{const.}
\end{equation}
Since the prediction is linear in $a_i$, the curvature and maximum of the likelihood give its standard deviation $\sigma_a$ and unconstrained amplitude estimate $\mu_a$:
\begin{align}
 \sigma_a^{-2}&=\bm s_i^T\bm\Sigma^{-1}\bm s_i,\notag\\
 \mu_a&=\sigma_a^2\bm s_i^T\bm\Sigma^{-1}(\bm d-\bm m_0).
\end{align}
We use a uniform prior on $a_i\ge0$. With $\Phi$ the standard normal cumulative distribution and $z_0=\Phi(-\mu_a/\sigma_a)$ the Gaussian probability below zero, the 95\% upper limit of the truncated posterior is
\begin{equation}
 a_{i,95}=\mu_a+\sigma_a\Phi^{-1}[z_0+0.95(1-z_0)].
\end{equation}
{Here $\Phi^{-1}$ is the standard normal quantile function, and the primordial-amplitude upper limit is $A_{t,i,95}=A_{t,\mathrm{ref}}a_{i,95}$. If $\bm s_i^T\bm\Sigma^{-1}\bm s_i=0$, this likelihood and uniform prior yield no finite upper limit. When low-multipole Planck data are included, we integrate the non-Gaussian posterior numerically. These data comprise TT and EE over $2\le\ell\le29$; low-multipole TE is excluded.}

{We derive the limits using the completed CLASS spectra for bins 1--20 and semi-analytic spectra for bins 21--25. Conditional limits for each wavenumber band are evaluated with each experiment's window functions, covariance, background and foreground model, and amplitude prior.}

\subsection{Conversion to gravitational-wave density}
\begingroup
With the normalization in Eq.~\eqref{eq:band}, the primordial-amplitude limit converts to the present band-averaged density as~\cite{WatanabeKomatsu,Kite}
\begin{equation}
 \langle\Omega_{\rm GW}\rangle_{k,95}h^2
 =\frac{A_{t,95}h^2}{24\Delta H_0^2}
 \langle [\partial_\eta T(k,\eta_0)]^2\rangle_k
 \label{eq:omega}
\end{equation}
The background quantities are those in Table~\ref{tab:parameters}. For bins 1--8, we obtain $[\partial_\eta T(k,\eta_0)]^2$ from the present-day CLASS metric transfer and integrate it directly over the band. This calculation includes tensor neutrino anisotropic stress and retains the oscillation phase within each band. For bins 9--16, we use the adiabatic-oscillator estimate for $u=aT$, $\overline{(\partial_\eta T)^2}\simeq(k^2+\mathcal H^2)(u^2+(\partial_\eta u)^2/k^2)/(2a^2)$, replacing $(\partial_\eta T)^2$ in Eq.~\eqref{eq:omega} by its oscillation average. For bins 17--25, the asymptotic solution for modes entering during radiation domination gives
\begin{equation}
 \langle\Omega_{\rm GW}\rangle_{k,95}h^2
 \simeq\frac{A_{t,95}D_\nu^2\Omega_r h^2}{48\Delta}
 \label{eq:omega-asymptotic}
\end{equation}
The latter two prescriptions agree to 1.1\% at bin 16. This quantity is a band average per logarithmic wavenumber; the integrated density is $\Delta$ times this average. We apply the same conversion coefficient in each band to TT, TE, EE, BB, and their combined upper limits. This conversion leaves the amplitude likelihoods and limits unchanged. For comparison, we also show bands that fail the short-wavelength condition $k_-/H_0\ge10$.
\endgroup

\section{Constraints and inter-experiment correlations}\label{sec:results}

\begin{figure}[tb]
\centering\includegraphics[width=.98\textwidth]{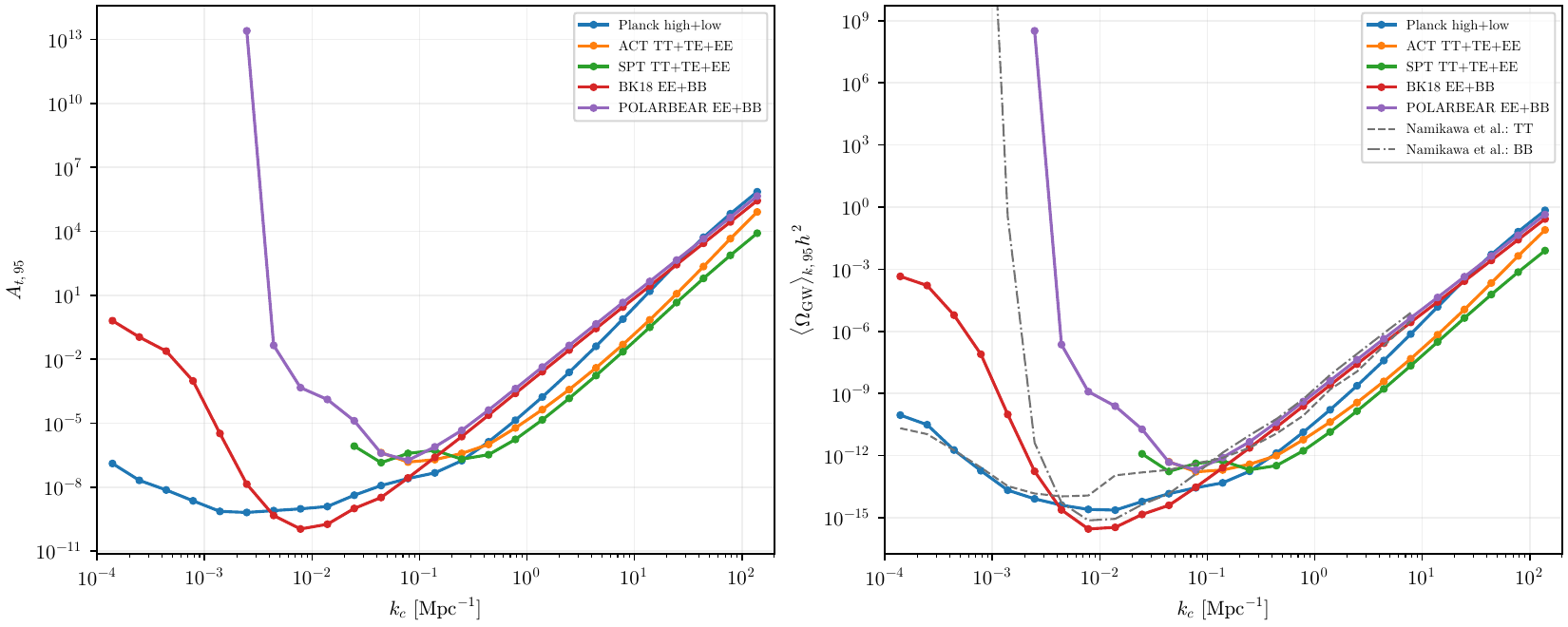}
\caption{{Conditional 95\% upper limits for each experiment. {Bins 1--20 use numerical CLASS spectra, while bins 21--25 use semi-analytic approximations to reduce the cost of high-wavenumber integration.} The left panel shows primordial band amplitudes and the right panel the band-averaged density in Eq.~\eqref{eq:omega}. The analyses use residuals relative to fixed reference models and the released within-experiment covariance. {POLARBEAR combines the EE and BB likelihoods assuming independence. EE marginalizes over foregrounds and calibration; BB uses a diagonal covariance approximation from the released errors. Bins 6--7 use EE alone because the BB response is unresolved.} Planck low-$T$, low-$E$, and high-multipole blocks are factorized. {Both panels include every band with a finite limit. The gray TT and BB reference curves} are the limits in Fig.~4 of Namikawa et al.~\cite{Namikawa2019}, multiplied by $h^2$ and retaining their published density-conversion convention.}}\label{fig:limits}
\end{figure}

{Figure~\ref{fig:limits} compares primordial-amplitude limits and their density conversion over the same wavenumber range. Finite limits are obtained for bins 1--25 with Planck and BK18, bins 11--25 with ACT, and bins 10--25 with SPT; all reach $k_c\simeq139\,\mathrm{Mpc}^{-1}$. The density curves are band averages rather than phase-dependent evaluations at the band center. Within the short-wavelength domain, the smallest density limits are $2.9\times10^{-16}$ for BK18 in bin 8 and $2.3\times10^{-15}$ for Planck in bin 9. The limits weaken at high wavenumber as the CMB response decreases, allowing the sensitivity of each observation to be compared throughout this range.}

{For POLARBEAR, we obtain finite EE limits for bins 6--25 and BB limits for bins 8--25. Figure~\ref{fig:limits} shows their combined upper limit under the independence approximation, with EE alone used for bins 6--7. The strongest combined band-averaged density limit is $2.0\times10^{-13}$ at $k_c\simeq0.0781\,\mathrm{Mpc}^{-1}$. Although weaker than BK18, this provides a constraint from a different observation.}

{The absence of finite limits for ACT bins 1--10 and SPT bins 1--9 reflects unresolved window-projected spectral tails rather than a fundamental absence of low-wavenumber sensitivity. The exact response can remain nonzero, but these bands are omitted from the constraint comparison because they provide no improvement in the present analysis. POLARBEAR EE bins 1--5 and BB bins 1--7 are omitted for the same reason.}

Only one primordial band is nonzero at a time. Since angular responses from different bands overlap, the envelope of all limits cannot be interpreted as an allowed simultaneous multiband spectrum. Comparisons with Namikawa et al.~\cite{Namikawa2019} must also account for differences in band width, data selection, and density conversion.

{Near $k\sim1\,\mathrm{Mpc}^{-1}$, SPT TT, TE, and EE give $\langle\Omega_{\rm GW}\rangle_{k,95}h^2=1.7\times10^{-12}$ at $k_c\simeq0.78\,\mathrm{Mpc}^{-1}$ (bin 16). At the same wavenumber, the published TT curve of Namikawa et al.~\cite{Namikawa2019}, displayed with the common factor $h^2$, gives $8.6\times10^{-11}$; our individual-experiment limit is tighter by a factor of about 50. In the adjacent band centered at $k_c\simeq1.39\,\mathrm{Mpc}^{-1}$ (bin 17), the SPT limit is $1.4\times10^{-11}$, a factor of about $1.1\times10^2$ tighter than the corresponding published limit of $1.6\times10^{-9}$. These comparisons include differences in data, spectra, likelihoods, and density-conversion conventions. The maximum band center also increases from approximately $7.8\,\mathrm{Mpc}^{-1}$ to $139\,\mathrm{Mpc}^{-1}$, extending the range of the earlier numerical analysis by a factor of about 18.}

\subsection{Inter-experiment correlations and combined limits}
\label{sec:cross-experiment}
\begingroup
Measurements made with different instruments can share the same CMB realization, lensing, and foreground emission over common sky regions and multipoles. Retaining within-experiment covariance and multiplying independent experiment likelihoods are therefore distinct assumptions. We compare the envelope of individual limits, the independent combination, and sensitivity to unknown correlations using common inputs.

The comparison includes five experiments: Planck, ACT, SPT, BK18, and POLARBEAR. We describe high-multipole Planck TT, TE, and EE, ACT TT, TE, and EE, SPT TT, TE, and EE, and BK18 EE and BB with Gaussian amplitude likelihoods obtained from residuals relative to fixed reference models and public covariance matrices. The Planck likelihood also includes the public non-Gaussian low-multipole TT (Commander) and EE (LoLLiPoP) likelihoods, while the POLARBEAR likelihood combines EE and BB as described in Sec.~\ref{sec:inference}. Each likelihood retains the within-experiment covariance treatment specified there. We assume independence between POLARBEAR and the other experiments, and factorize the low-$T$ and low-$E$ blocks relative to the other blocks. The correlation coefficients below apply to the Gaussian amplitude estimates.

For each wavenumber band, we denote the unconstrained maximum-likelihood amplitude of Gaussian input $i$ by $\widehat A_i$ and its standard deviation by $\sigma_i$, and form $\boldsymbol a=(\widehat A_1,\ldots,\widehat A_n)^T$. These estimates incorporate the known within-experiment covariance. For an unknown inter-experiment covariance $C$, the compressed likelihood is
\begin{equation}
 -2\ln\mathcal L_{\rm G}(A_t;C)=(\boldsymbol a-A_t\boldsymbol 1)^TC^{-1}(\boldsymbol a-A_t\boldsymbol 1)+\mathrm{const.}
\end{equation}
Here $\boldsymbol 1$ is the vector of ones and $n$ is the number of Gaussian inputs with a finite numerical response in the band. Inverse-covariance weighting gives the combined mean,
\begin{equation}
 \widehat A(C)=\frac{\boldsymbol 1^TC^{-1}\boldsymbol a}{\boldsymbol 1^TC^{-1}\boldsymbol 1}.
\end{equation}
The variance of the combined estimate is
\begin{equation}
 \sigma_A^2(C)=(\boldsymbol 1^TC^{-1}\boldsymbol 1)^{-1}.
\end{equation}
These mean and variance describe the Gaussian block. We denote the five experiments by $\mathcal E=\{\mathrm{Planck},\mathrm{ACT},\mathrm{SPT},\mathrm{BK18},\mathrm{PB}\}$ and their individual likelihoods by $\mathcal L_e$. With $C_0=\operatorname{diag}(\sigma_1^2,\ldots,\sigma_n^2)$, the full likelihood is
\begin{equation}
 \mathcal L_{\rm all}(A_t;C)=
 \frac{\mathcal L_{\rm G}(A_t;C)}{\mathcal L_{\rm G}(A_t;C_0)}
 \prod_{e\in\mathcal E}\mathcal L_e(A_t).
 \label{eq:combined-experiments}
\end{equation}
Each $\mathcal L_e$ is the experiment likelihood specified in Sec.~\ref{sec:inference}, including the low-multipole contributions for Planck. The ratio replaces the independent Gaussian product by its correlated likelihood. Under independence, $C=C_0$ and the ratio is unity, leaving the product of all five experiment likelihoods. Within-experiment covariances remain included under this assumption.

We apply the uniform prior $A_t\geq0$ once and numerically integrate the non-Gaussian posterior to obtain its 95\% quantile, $U(C)$. The Planck-only analysis includes the same low-multipole likelihoods and reproduces the individual limits within 0.004\%. Halving the integration grid changes the combined limits by less than 0.023\%.

The combined curves below include all five experiments. Adding POLARBEAR changes the limits by at most 0.38\% across the 25 bands.

The envelope of individual limits is $\min_i U_i$. It is a comparison of observed individual upper limits, distinct from the 95\% quantile of a combined posterior. Independence corresponds to $C_{ij}=\sigma_i^2\delta_{ij}$. To examine correlation sensitivity, we use
\begin{equation}
 C_{ij}(\rho)=\sigma_i\sigma_j[(1-\rho)\delta_{ij}+\rho],\quad 0\leq\rho\leq\rho_{\max}<1.
 \label{eq:common-correlation}
\end{equation}
This covariance family is positive definite over the specified interval. The coefficient $\rho$ describes correlations between compressed amplitude estimates, rather than the sky-overlap fraction itself. We use $\rho_{\max}=0.25,0.5,0.75,0.95$ as sensitivity-test settings, rather than estimates inferred from the observations. Within each allowed interval, we numerically evaluate
\begin{equation}
 U_{\rm family}(\rho_{\max})=\sup_{0\leq\rho\leq\rho_{\max}}U[C(\rho)].
\end{equation}
Figure~\ref{fig:corr-family} shows the cases 0.5 and 0.95. This procedure gives a conservative credible upper limit within the specified one-parameter family. Its scope is limited to that family; it provides neither a bound over all positive-definite covariance matrices nor a guarantee of 95\% frequentist coverage.

Adding independent inputs can shift the observed maximum-likelihood amplitude, yielding a 95\% upper limit weaker than the strongest individual limit. The relative ordering of the envelope and combined limit therefore depends on more than variance reduction. Maximizing the correlation also differs from maximizing the upper limit, because correlations change the combined best fit and weights as well as the variance~\cite{ValassiChierici2014}. Across all 25 bands, maximizing over $0\leq\rho\leq0.95$ gives upper limits 1.0--1.5 times those obtained under independence. The maximizing $\rho$ varies between bands and can lie within the interval. This variation illustrates the sensitivity to assumed correlations rather than measuring an uncertainty range for the actual inter-experiment covariance.

The strongest five-experiment limit under independence is $\langle\Omega_{\rm GW}\rangle_{k,95}h^2=2.9\times10^{-16}$ at $k_c\simeq7.8\times10^{-3}\,\mathrm{Mpc}^{-1}$. BK18 dominates this band; the ratio of the published BB limit of Namikawa et al., $7.5\times10^{-16}$, to our result is about 2.6. At $k_c\simeq0.78\,\mathrm{Mpc}^{-1}$, the independent combination reaches $1.6\times10^{-12}$, a factor of about 52 tighter than the published TT limit at the same wavenumber. The limits maximized within the specified correlation family are $3.3\times10^{-16}$ and $2.3\times10^{-12}$, respectively. These comparisons include the differences in data, likelihoods, and density-conversion conventions discussed above.

\begin{figure}[tb]
\centering\includegraphics[width=.98\textwidth]{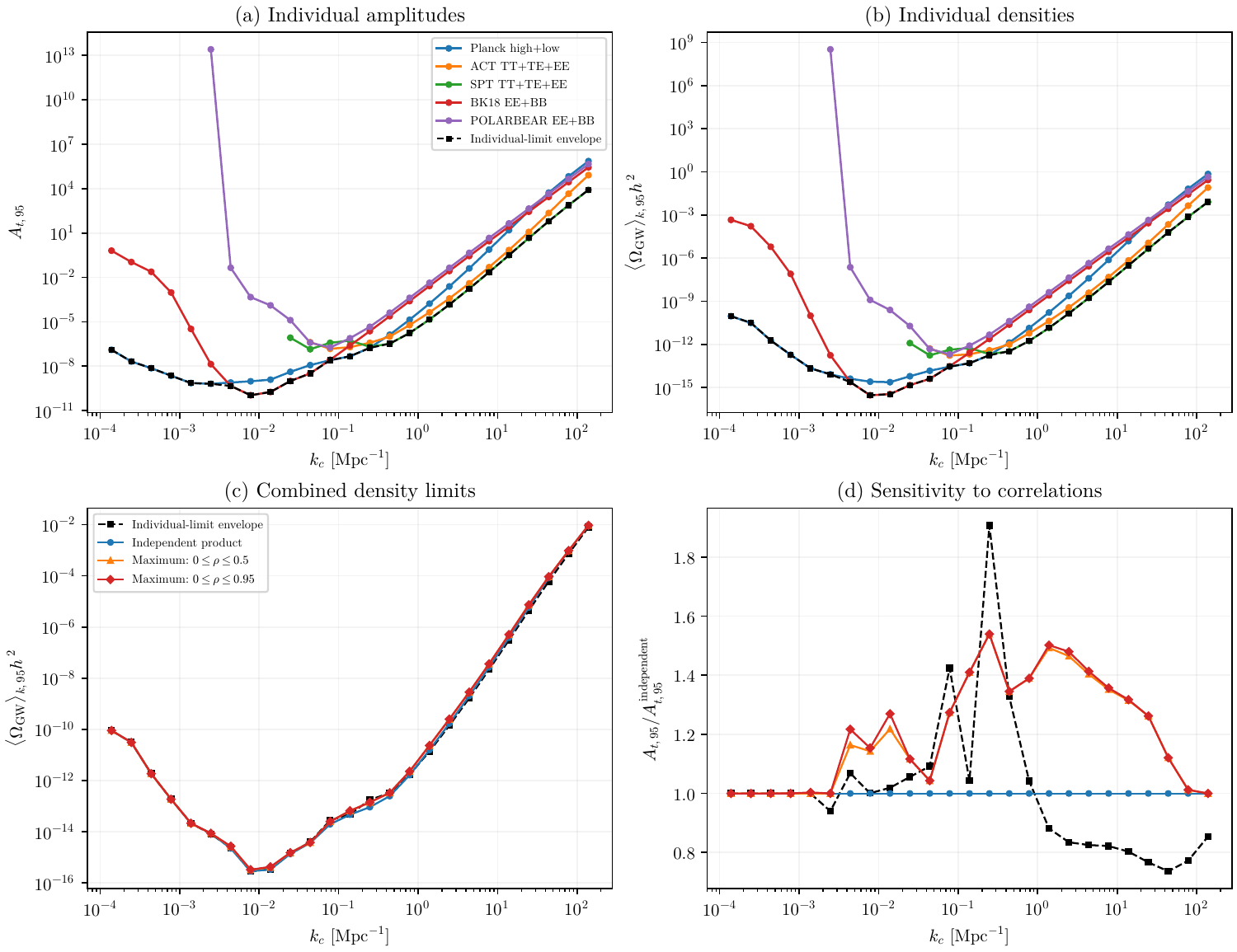}
\caption{{Individual limits and correlation-dependent combinations for high- and low-multipole Planck, ACT, SPT, BK18, and POLARBEAR. Panels (a) and (b) show individual amplitude and density limits and their envelopes; (c) compares the envelope, the independent product, and the maximum upper limit within the specified correlation family; (d) gives ratios relative to independence. Points mark results for each wavenumber band and lines connect them. All 25 bands are shown. We vary $\rho$ among the Gaussian amplitudes of high-multipole Planck, ACT, SPT, and BK18. The envelope is the minimum of individual upper limits, distinct from the quantile of a combined posterior. The family maximum applies to the specified covariance family.}}\label{fig:corr-individual}\label{fig:corr-independent}\label{fig:corr-family}
\end{figure}

A full combination of the observations requires constructing the shared CMB, foreground, and calibration covariance from the observed sky regions and window functions, and defining its allowed set from the associated uncertainties. This modeling must also include the cross-block correlations involving POLARBEAR and the low-multipole blocks for which independence is assumed here. Marginalizing over a prior on $\rho$ and maximizing the upper limit over its allowed range address different questions and should be reported separately.
\endgroup

\section{{Discussion}}\label{sec:discussion}
We have constructed the five responses to a finite-width primordial tensor spectrum from common transfer amplitudes and a common band normalization. The polarization scaling $D_\ell\propto\ell(\ell+1)$ corresponds to cancellation between the source and projection phases; the residual phase across the finite width controls its breakdown at high multipoles. TE requires the relative temperature--polarization phase, since the geometric mean of the auto-spectra alone leaves its sign undetermined. For TT and curl, the line-of-sight integral must be retained at the amplitude level rather than extending the closed observer-local limit to all distances.

{Figure~4 and the discussion in Namikawa et al.~\cite{Namikawa2019} show that curl gives weaker density constraints than temperature and $B$ modes while providing complementary information. Motivated by this comparison, our observational analysis focuses on TT, TE, EE, and BB. We nevertheless present the curl transfer function and semi-analytic spectrum because curl probes metric perturbations along the full photon path and is a complementary observable.}

\subsection{{Extension to high wavenumber}}
\begingroup
Ng~\cite{Ng2022} evaluated temperature angular spectra from the line-of-sight tensor Sachs--Wolfe (SW) integral [Eqs.~(18)--(21) of that paper] and derived limits up to $k\simeq140\,\mathrm{Mpc}^{-1}$, including $k\sim10$--$100\,\mathrm{Mpc}^{-1}$. The limits in Fig.~2 of that study follow from requiring the tensor contribution to remain below the observed temperature power at representative multipoles [its Eq.~(32) and Sec.~IV C].

We develop the analysis of this high-wavenumber regime into a temperature and polarization likelihood analysis using the full spectra over the multipole ranges included in the data. Our semi-analytic response includes the finite width of recombination and photon scattering sources, and applies to TE, EE, and BB as well as TT. Projecting these spectra through the observational window functions and combining them with the public covariance matrices allows each primordial band amplitude to be evaluated through the 95\% quantile of its posterior. We thus obtain statistically defined 95\% credible upper limits, conditional on the adopted reference models and priors, up to $k_c\simeq139\,\mathrm{Mpc}^{-1}$. Figure~\ref{fig:corr-family} compares their dependence on inter-experiment correlation assumptions, including the non-Gaussian low-multipole Planck likelihoods.
\endgroup

\begin{figure}[tb]
\centering\includegraphics[width=.85\textwidth]{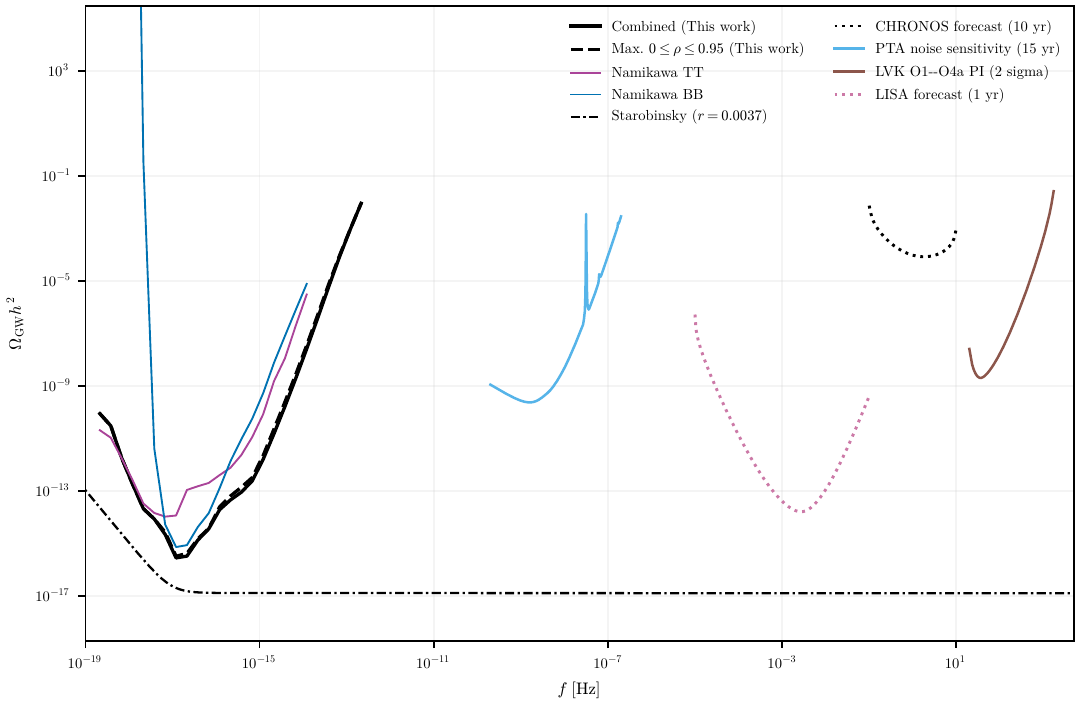}
\caption{{Combined frequency-domain upper limits compared with previous constraints and observational sensitivities. The black solid and dashed curves (This work) show our five-experiment combination under independence and the maximum within the specified common-correlation family, respectively. POLARBEAR enters as an independent likelihood, and all calculated bands are shown. The Namikawa et al. curves retain their published conversion convention. The black dot-dashed curve is the Starobinsky reference model with $r=0.0037$ used in the BB--EE study. PTA denotes the effective noise sensitivity of the 15-year data, LVK the O1--O4a $2\sigma$ PI sensitivity, and LISA and CHRONOS the one- and ten-year forecasts, respectively. The statistical meaning of each curve is described in the text.}}
\label{fig:frequency}
\end{figure}

\subsection{{Frequency-domain comparison}}
\begingroup
Figure~\ref{fig:frequency} converts the combined limits in Fig.~\ref{fig:corr-family} to frequency using $f=ck/(2\pi)$ with $a_0=1$. We show the product of the five experiment likelihoods under independence and the limit maximized within the common-correlation family $0\le\rho\le0.95$, with POLARBEAR included as an independent likelihood. The final band center, $k_c\simeq139\,\mathrm{Mpc}^{-1}$, corresponds to $f_c\simeq2.15\times10^{-13}\,\mathrm{Hz}$. The temperature and $B$-mode limits of Namikawa et al.~\cite{Namikawa2019} are overplotted to illustrate the low-frequency range probed by the CMB. Their curves retain the published density-conversion convention, so differences between curves include this convention as well as band width and data selection.

For theoretical comparison, we use the same Starobinsky reference model~\cite{Starobinsky1980} as in the BB--EE study~\cite{InoueEEBB}. Including the smooth, oscillation-averaged transfer through matter--radiation equality, its present energy-density spectrum is~\cite{Clarke}
\begin{equation}
 \begin{aligned}
 \Omega_{\rm GW}^{\rm St}(f)h^2\simeq{}&
 \frac{3rA_s\Omega_r h^2}{128}
 \left(\frac{f}{f_*}\right)^{n_t}\\
 &{}\times\left[\frac12\left(\frac{f_{\rm eq}}{f}\right)^2
 +\frac{16}{9}\right].
 \end{aligned}
 \label{eq:starobinsky-reference}
\end{equation}
We adopt $r=0.0037$, $n_t=-r/8$, $A_s=2.1\times10^{-9}$, $h=0.674$, and $\Omega_r=9.0\times10^{-5}$. Here $r$ is the tensor-to-scalar ratio at $k_*=0.05\,\mathrm{Mpc}^{-1}$ and $n_t$ is the primordial tensor spectral index. The pivot frequency is $f_*=ck_*/(2\pi)\simeq7.7\times10^{-17}\,\mathrm{Hz}$, and we use the equality frequency $f_{\rm eq}=1.7\times10^{-17}\,\mathrm{Hz}$. The bracket describes the transfer effect associated with matter--radiation equality and produces the knee between the low-frequency scaling $f^{n_t-2}$ and the high-frequency scaling $f^{n_t}$. Equation~\eqref{eq:starobinsky-reference} specifies the broadband reference curve shown in Fig.~\ref{fig:frequency}; its display near the present horizon is a formal continuation of this approximation.

We also show the effective PTA noise sensitivity~\cite{NANOGravNoise2023}, the LVK O1--O4a power-law-integrated (PI) sensitivity~\cite{LVKO4a2025}, and the projected LISA and CHRONOS sensitivities~\cite{Schmitz2021,CHRONOSCQG2026,Thrane2013} used in the BB--EE study~\cite{InoueEEBB}. The LISA and CHRONOS forecasts assume one and ten years of observation, respectively, known noise, and a signal-to-noise ratio of unity. CHRONOS uses the same instrument configuration as the BB--EE study. The label ``This work'' identifies our combination of current data. These sensitivity curves and the CMB bandwise 95\% upper limits are different statistical quantities; the figure compares their complementary frequency ranges.
\endgroup

\section{{Conclusions}}\label{sec:conclusion}
\begingroup
We have generalized the semi-analytic EE and BB formalism for finite wavenumber bands to TT, TE, and the deflection curl. Comparisons with direct CLASS calculations quantify the approximation accuracy. Combining the direct and semi-analytic calculations provides 25 bands extending to $k_c\simeq139\,\mathrm{Mpc}^{-1}$, a factor of about 18 above the maximum band center of approximately $7.8\,\mathrm{Mpc}^{-1}$ in the numerical analysis of Namikawa et al.~\cite{Namikawa2019}.

Temperature and polarization analyses retaining the public covariance matrices yield 95\% credible upper limits for each primordial band. The strongest band-averaged density limit from the five-experiment combination under independence is $\langle\Omega_{\rm GW}\rangle_{k,95}h^2=2.9\times10^{-16}$ at $k_c\simeq7.8\times10^{-3}\,\mathrm{Mpc}^{-1}$, a factor of about 2.6 tighter than the published BB limit of Namikawa et al. Near $k\sim1\,\mathrm{Mpc}^{-1}$, we reach $1.6\times10^{-12}$ at $k_c\simeq0.78\,\mathrm{Mpc}^{-1}$, a factor of about 52 tighter than their published TT limit of $8.6\times10^{-11}$. These comparisons include differences in data, likelihood settings, and density-conversion conventions. Each upper limit corresponds to a model that adds a signal in one primordial band to a fixed reference cosmology; for POLARBEAR EE, foreground and calibration parameters are marginalized following the public likelihood.

For the five-experiment combination, we compare independence with upper limits maximized over common correlations $0\le\rho\le0.95$ between the Gaussian amplitude estimates. POLARBEAR is included as an independent likelihood. The maximized limits are 1.0--1.5 times the independent limits, quantifying the effect of unknown inter-experiment correlations within the specified family. These results establish the temperature and polarization sensitivity to localized primordial gravitational waves up to high wavenumber and illustrate their complementarity with gravitational-wave observations at other frequencies.
\endgroup

\begin{acknowledgments}

We thank Y.~Mikura, T.~Okumura, Y.~Chinone, T.~Namikawa, and G.~Galloni for helpful discussions and comments. Y.I. acknowledges support from NSTC and CHiP in Taiwan under Grant No.~115-2112-M-008-022-. M.H. is partially supported by the Yushan Fellow Program of the Ministry of Education (MOE), Taiwan.
\end{acknowledgments}

\section*{{Data Availability}}
\begingroup
The observational data and associated likelihood files used in this work are publicly available. BK18, ACT DR6, and POLARBEAR data are distributed through NASA LAMBDA~\cite{BK18Data,ACTDR6Data,POLARBEAREEData,POLARBEARBBData}; Planck PR4 HiLLiPoP and LoLLiPoP data are available from NERSC~\cite{PlanckPR4Data}; Planck 2018 low-multipole TT data are provided through the Cobaya distribution~\cite{PlanckLowTData}; and SPT-3G D1 data are available from the official SPT data repository~\cite{SPTD1Data}. We use the public software CLASS~\cite{CLASS}, Cobaya~\cite{Cobaya}, and candl~\cite{candl}. The combined upper-limit curves for the five experiments and the individual-experiment results obtained in this work will be released on the CHRONOS website (\url{https://chronos.phy.ncu.edu.tw}). The narrow-band spectra, modified CLASS source, and analysis code are available from the corresponding author upon reasonable request.
\endgroup

\appendix
\section{{Radial kernels regular at the turning point}}\label{app:radial}
{We evaluate the radial kernel near the turning point using a uniform asymptotic expansion of the spherical Bessel function.}
For $\ell<100$, we evaluate the spherical Bessel function directly; for $\ell\ge100$, we use the leading uniform asymptotic term~\cite{DLMF}. With $\nu=\ell+1/2$ and $z=x/\nu$, this gives
\begin{equation}
j_\ell(x)\simeq\sqrt{\frac\pi{2x}}\left(\frac{4\zeta}{1-z^2}\right)^{1/4}
\frac{\operatorname{Ai}(\nu^{2/3}\zeta)}{\nu^{1/3}},
\end{equation}
The turning-point variable is defined on either side of $z=1$ by
\begin{equation}
 \frac23|\zeta|^{3/2}=\begin{cases}
 \displaystyle\ln\frac{1+\sqrt{1-z^2}}{z}-\sqrt{1-z^2},&0<z\le1,\\[3pt]
 \sqrt{z^2-1}-\arccos(1/z),&z\ge1,
 \end{cases}
\end{equation}
We take $\zeta\ge0$ below the turning point and $\zeta\le0$ above it.
Here $\operatorname{Ai}$ is the Airy function and $\zeta$ is the turning-point variable defined above. As $z\to1$, we use $[4\zeta/(1-z^2)]^{1/4}\to2^{1/3}$. The kernel is exact for $\ell<100$, while the source retains its high-$k$ approximation.
\bibliography{references}
\end{document}